\documentstyle[12pt,epsf,aps]{revtex}
\begin{document}
\baselineskip 0.3in
\title{A Microscopic Mechanism for Muscle's Motion}
\author{{AI Bao-Quan, WANG Xian-Ju, LIU Guo-Tao and  LIU Liang-Gang \footnote{Email: stdp05@zsu. edu. cn}}\\
{Department of Physics, ZhongShan University, GuangZhou, P. R. China.}\\
{Nakano M}\\
{Department of information science, University of Occupational and Environmental Health, Japan.}\\
{Matsuura H}\\
{Department of Project Center, National Graduate Institute for Policy Studies, Japan.}}

\parskip 0pt
\maketitle
\begin{abstract}
\baselineskip 0.3in

\indent The SIRM (Stochastic Inclined Rods Model) proposed by H. Matsuura and M. Nakano can explain the muscle's motion perfectly, but the intermolecular potential between myosin head and G-actin is too simple and only repulsive potential is considered. In this paper we study the SIRM with different complex potential and discuss the effect of the spring on the system. The calculation results show that the spring, the effective radius of the G-actin and the intermolecular potential play key roles in the motion. The sliding speed is about $4.7\times10^{-6}m/s$ calculated from the model which well agrees with the experimental data.
Pacs: 87. 10. +e, 87. 14. Gg, 05. 90. +m.
\end{abstract}

\vskip 0.3in
\baselineskip 0.3in
\section {Introduction}
\parskip 0pt
\indent
 The great progress have been made on the observations of molecular motor
 since H. F. Huxley and  H. E. Huxley \cite{1}\cite{2}\cite{3} proposed the rotating cross-bridge model to study muscular mobility in 1954. More and more biologists and theoretical biophysicists concentrate on studying the mechanism of the living system and mincing the biological structure to invent subtle artificial instrument.
Based on different system, three different families of motor proteins have
been identified \cite{4}\cite{5}\cite{6}, Kinesins and dyneins move along
tubulin filaments\cite{7}, myosin move along actin filaments \cite{8}\cite{9}.The motion mechanism of the three kinds of motors can be described as follows \cite{4}
\cite{10}\cite{11}: firstly, the motor binds Adenosinetriphosphate (ATP),
then hydrolyzes the bound ATP and absorbs the energy, subsequently, the motor releases the products Adenosinediphosphate (ADP) and Phosphate (P), so the motor move constantly within the chemical cycle. The molecular motors play a key role in transuding chemical energy into mechanical work at a molecular scale. Up to date there have been three kinds of archetype models proposed to study the molecular motor system: namely the fluctuation models, the phase transition models and the resonance models. \\
\indent
\parskip 0pt
One archetype fluctuation model is ratchets model. This model looks molecular
motors as heavily damped Brownian particles and the model explains the
molecular motor's motion by studying a single particle in a potential inspired
by Feynman's ratchet. M. O. Magnasco \cite{12} and R. D. Astumian\cite{13}\cite{25} have
proposed the forced thermal ratchets model, the fluctuant force break the symmetry of the system, so that the system can absorb the energy from the surroundings, the motor undergoes similar Brownian motion and produces net transport. Although the fluctuation models have discussed different aspects of molecular motors, there still exits a matter of debate on the mechanism of its motion.\\
\indent
\parskip 0pt
The second typical model is phase transition model: the two state model proposed by F.Julicher\cite{4} and isothermal ratchet model proposed by A. Pameggiani\cite{14}. Using the basic idea of the Ising model for phase transitions, they have studied a single motor or many motors moving in asymmetry periodic potential. In these models the microtubule and molecular motor are treated as periodic polar railway tracks and
carnot engines with cargoes, the motor consumes chemical energy and causes the system conformation change, so the phase transition occurs between two states. The models give a good explanation on the efficiency of the motor.\\
\indent
\parskip 0pt
The newly type resonance model is named the Stochastic Inclined Rods Model (SIRM)
proposed by H. Matsuura and H. Matsuura\cite{15-20}, in the model  the energy of motion was supplied from the random noise and the system always moves to one direction by using stochastic resonance. The movement of the system does not break the second law of thermodynamics, because the actin-myosin system is open to the surroundings and the energy flows in from their surroundings. The SIRM presents a perfectly sliding mechanism of the actin-myosin system, but the interaction is too simple and only repulsive force is considered. The main aim of our paper is that based on SIRM we study muscle's motion under a complex intermolecular potential and discuss the effect of the spring on the system.
\parskip 0pt

\section {Model and Formalism}
\indent In order to describe the motion of actin-myosin system, we simulate it by a mechanical model as shown in Fig. 1. SIRM is composed of three parts: an inclined spring, a myosin head and myosin bundle. Our model proposed a microscopic mechanism for the actin-myosin system: The random noise is from the water molecules generated through the heat energy of ATP hydrolysis. ATP supplies the heat energy and provides heat energy to the surrounding water molecules, which accelerates the movement of the water molecules, So the random noise (the water molecules) can interact with the myosin head, and the resonance occurs between the myosin head and the noise, then it collides with a G-actin and obliquely kicks the G-actin sphere, because the direction of vibration is inclined against the line of the actin fibers, myosin molecules obtain the propellant force along the direction of the fiber and the myosin can move to one direction.\\
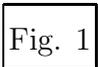
\begin{center}
\begin{tabular}{|c|}\hline
Fig. 1\\\hline
\end{tabular}
\end{center}
\indent Let us set up the dynamic equations of the system.
As for the intermolecular potential between myosin head and G-actin we adopt Jennard-Jones potential:
\begin{equation}
\label{e1}
U_a={\sum\limits_{i=0}^{i=n}U_a^0(pe^{-r_{i}/jk}-qr_{i}^{-6})}.
\end{equation}
\indent Where $r_{i}=\sqrt{(x-x_{i})^{2}+(y-y_{i})^{2}}-R$,  $(x_{i},y_{i})$ means a center of i-th G-actin glove, and R is its effective radius. $p, q,U_a^0,jk$ are the parameters of the potential.\\
The potential of the myosin rod is approximately expressed as follows:\\
\begin{eqnarray}
\label{e2}
U_s=K_{l}exp(-\sqrt{(x-x_{2})^{2}+(y-y_{2})^{2}}+L)\nonumber\\
\nonumber  +K_{l}exp(\sqrt{(x-x_{2})^{2}+(y-y_{2})^{2}}-L)\nonumber\\
 +{1\over 2}K_{\theta}(\theta-\theta_{0})^{2}.
\end{eqnarray}
\indent
Where $K_{\theta},K_{l}$ are tangent constant and radial constant of the spring,
 and $L$ is  initial length of the spring , ${\tan(\theta)}={(x_{2}-x)\over y}, \theta,{\theta}_{0}$ are current angle and initial angle between the rod and level axis.
$ (x_{2},y_{2})$is the center of gravity of myosin bundle.\\
\indent
The equations of the myosin head are:

\begin{equation}
\label{e3}
m{\frac{\partial^{2} x}{\partial t^{2}}=-{\frac{\partial (U_{a}+U_{s})}
{\partial x}}+F_{x}(t)-\alpha{\frac{\partial x}{\partial t}}}.
\end{equation}
\begin{equation}
\label{e4}
m{\frac{\partial^{2} y}{\partial t^{2}}=-{\frac{\partial (U_{a}+U_{s})}
{\partial y}}+F_{y}(t)-\beta{\frac{\partial y}{\partial t}}}.
\end{equation}
\indent With regard to the center of gravity $(x_{2}, y_{2})$ of the filament, we set a similar equation:

\begin{equation}
\label{e5}
M{\frac{\partial^{2} x_{2}}{\partial t^{2}}=-{\frac{\partial U_{s}}
{\partial x_{2}}}+F_{x_{2}}(t)-\eta{\frac{\partial x_{2}}{\partial t}}}.
\end{equation}
The variable $y_{2}$ is fixed since the myosin filament does not significantly move along the $y$-direction compared to the $x$-direction.\\

\indent
Where  $\alpha, \beta, \eta $ are viscous constants, $M$ is the mass of the myosin bundle and $m$ is the mass of the myosin head. $F_{x}(t), F_{y}(t),F_{x_{2}}(t)$ are fluctuation of the thermal noise, the noise is Gauss white noise and it satisfies the following fluctuation-dissipative relations\cite{21}\cite{22}\cite{23}.\\

\begin{equation}
\label{e6}
<F_{a}(t)>=0.
\end{equation}
\begin{equation}
\label{e7}
<F_{a}(t)F_{b}(s)>=2k_{B}T{\zeta}{\delta}_{a,b}{\delta}(t-s).
\end{equation}
where $a,b=x,y,x_{2};  {\zeta}={\alpha},{\beta},{\eta};  k_{B}$ is Boltzmann constant.  $T$ is absolute temperature, $t,s$ are time.

\section{Calculation results}
\indent
In order to solve the above equations we adopt a numerical method. The parameters of the equations are shown in the table.\\
\begin {center}
Table. 1. The Parameters of the Equations.\\
\end{center}
\tabcolsep 0.15in
\begin{tabular}{|c|c|c|c|c|c|}\hline
1 unit time &$10^{-4}$s& $\alpha, \beta, \eta$ &1& $U_{a}^{0}$&15000\\\hline
1 unit length &  $10^{-9}$m &  $ p, q$&100& $m$ &80 \\\hline
 1 unit mass& $1.66\times10^{-24}$kg &$L$ &10& $M$&400 \\\hline
$K_{\theta}$ & 450000& $K_{l}$ & 4500 &$\theta_{0}$ &$45^{\circ}$\\\hline
\end{tabular}\\

We solve the equations(1)-(7) with a numerical method and the numerical results are shown in Fig.2-Fig.6.\\
\begin{center}
\begin{tabular}{|c|}\hline
Fig. 2\\\hline
\end{tabular}
\end{center}
\indent The dashed line shows the position of myosin filament and the solid line describes the displacement of myosin head. From the figure we can know that the head and the filament, bound by the spring, move forward along the same direction as a whole. The movement of the head is a translational one as a whole, though the trace has irregularity and randomness like Brownian particles. We can know from the figure the average gliding speed of the myosin bundle is $4.7\times10^{-6}m/s$, On the other hand, the experimental data by Yanagida et al. is that the maximum gliding speed is $7\times10^{-6}m/s$. It is impressive that our prediction well agree with the experimental data.\\
\begin{center}
\begin{tabular}{|c|}\hline
Fig. 3\\\hline
\end{tabular}
\end{center}
\indent The solid line shows the stretching motion of the myosin rod, the vertical displacement of myosin head is described  with the dashed line. The figure shows that the collision  between the G-actin and the myosin head distorts their relative motion from a trigonometric function. This trace of the motion is given by an accurate trigonometric function if there are no interactions or the collisions. Truly the strain of the trigonometric function indicates that the myosin head kicks obliquely the G-actin when the spring is stretched by the resonance vibration.\\

\begin{center}
\begin{tabular}{|c|}\hline
Fig. 4\\ \hline
\end{tabular}
\end{center}
 \indent From the figure we can know that the constant of the spring play a key role in the motion, if the constant is zero, the system can not move. And if the constant increases, the velocity of the SIRM will increase rapidly.\\
\begin{center}
\begin{tabular}{|c|}\hline
Fig. 5\\ \hline
\end{tabular}
\end{center}
\indent  The figure can show that the effective radius of the G-actin is important for the system to move. If the G-actin is flat ($R=0$) the SIRM cannot move, on the other hand, if the effective radius of the G-actin is too big ($R=4nm$) the system cannot move, either. The system
can move at the highest speed at $ R=2.5nm$.\\
\begin{center}
\begin{tabular}{|c|}\hline
Fig. 6\\ \hline
\end{tabular}
\end{center}
\indent From the figure we can know that the repulsive potential can accelerate the system's
motion and the potential between the G-actin and myosin can decide the
motion's velocity.\\

\section{Summary and conclusion }
\indent Based on the SIRM we study the movement of actin-myosin system with a Jennard-Jones
potential. From the numerical results we can know there is relative sliding motion between the filament and the G-actin and the average gliding speed is $4.7\times10^{-6}m/s$ which well agree with the experimental data. When the system absorbs the energy from the thermal noise constantly through stochastic resonance, the intermolecular potential and the inclined rod make the filament move to one direction. The system can convert the random noise to one directional motion. The effective radius of G-actin, the constant of the spring and the intermolecular potential are important for the system to move. If the constant of the spring is small, the system cannot move, and if the effective radius of G-actin is too small or too big the system cannot move, either. The repulsive potential makes the system move fast. From the results we can predict the state of the muscle: if the constant of the spring
or the effective radius of G-actin become small, the SIRM can not move at all and muscle is fatigue, if the effective radius of G-actin is too big, the velocity of the system is zero and the actin-myosin system may be breakdown.\\
\indent SIRM is thermally open to the outer surroundings and it has a outer heat source of ATP. The heat or energy consumed by SIRM is directly supplied from the surroundings (i.e. the thermal motions of water molecules). The entropy always increases since the flow is positive, therefore, the system does not break the second law of thermodynamics. There have been many discussions on the movement of ratchets in a heat bath. Among them, we only consider one famous discussion of Feynman\cite{24}. Feynman showed from statistical discussions that the ratchets under an isothermal temperature cannot turn as a total since the ratchets turn to both sides with equal probability. However, he also showed in the same section, 46-2, of Ref.\cite{24} that it can turn to one side if a load is suspended from the one side even in an isothermal bath. Our model can work even under isothermal temperature conditions, since it has a  load, which is an intermolecular potential of the G-actin fastened to a Z-membrane.\\
\indent
Noise in dynamical systems is usually considered a nuisance. But in certain nonlinear systems including electronic circuits and biological sensory apparatus the presence of noise can in fact enhance the detection of the weak signals. From the above results, we point out the possibility that the actin-myosin system can use thermal noise around for the movement in similar way. This viewpoint is completely new as the traditional viewpoint thinks the water disturbs the motion as viscosity.

\newpage
\section{Figure and table captions}
\baselineskip 0.3in
{\bf{Table. 1.}} The parameters of the equations. $\alpha, \beta, \eta$ are viscous constants, $M, m$ are the mass of myosin bundle and myosin head , respectively, $L$ is initial length of the spring, $p, q,U_a^0,jk$ are the parameters of the potential, $K_{\theta},K_{l}$ are tangent constant and radial constant of the spring. \\

{\bf{Fig. 1.}}The model for actin-myosin system. The coordinates of the head of myosin and the end of myosin connected to the filament are $(x, y)$ and $(x_{2}, y_{2})$, respectively. The body of the myosin is assumed to be represented as  a spring.\\
{\bf{Fig. 2.}} The dashed line shows the position of myosin bundle and the solid line describes the displacement of the myosin head. The parameters are shown in the figures also.\\
{\bf{Fig. 3.}} The dash line shows the vertical movement of the myosin head and the solid line presents the stretching displacement $s$ of the rod and $s=\sqrt{(x-x_{2})^{2}+(y-y_{2})^{2}}-L$.\\
{\bf{Fig. 4. }}The relation between the constant of the spring and the x-displacement of the myosin bundle is shown in the figure. From the up to the bottom, different lives correspond to $K_{l}$: 5000,3000,1000,0, respectively. \\
{\bf{Fig. 5.}} The relation between the effective radius of the G-actin and the x-displacement of the myosin bundle is shown in the figure. From the up to the bottom, different lives correspond to $R$: 2.5, 1, 4, 0, respectively.\\
{\bf{Fig.6.}} The figure indicates the relationship of the intermolecular potential and the x-displacement of the myosin bundle. From the up to the bottom, different lives correspond to $jk$: 4, 3, 2, 1, respectively.\\
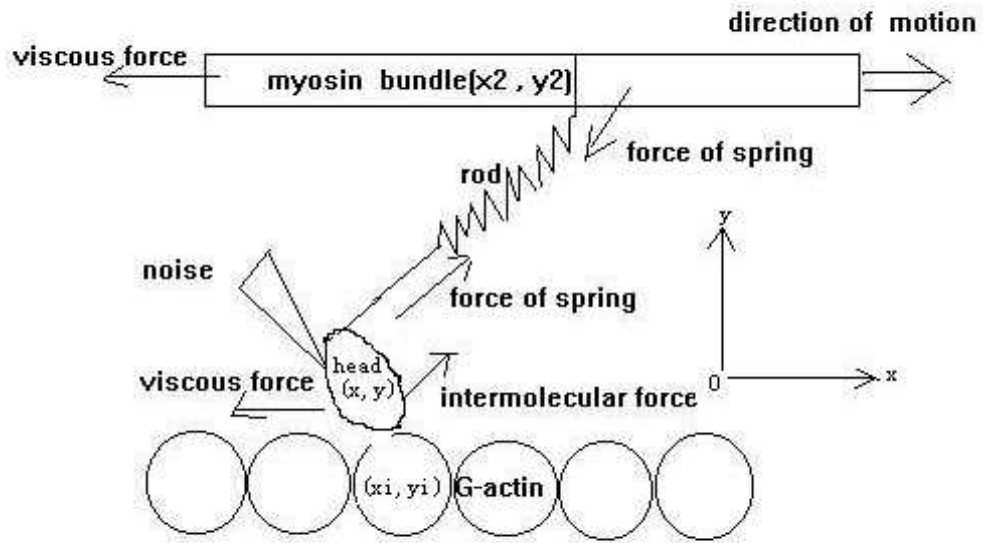
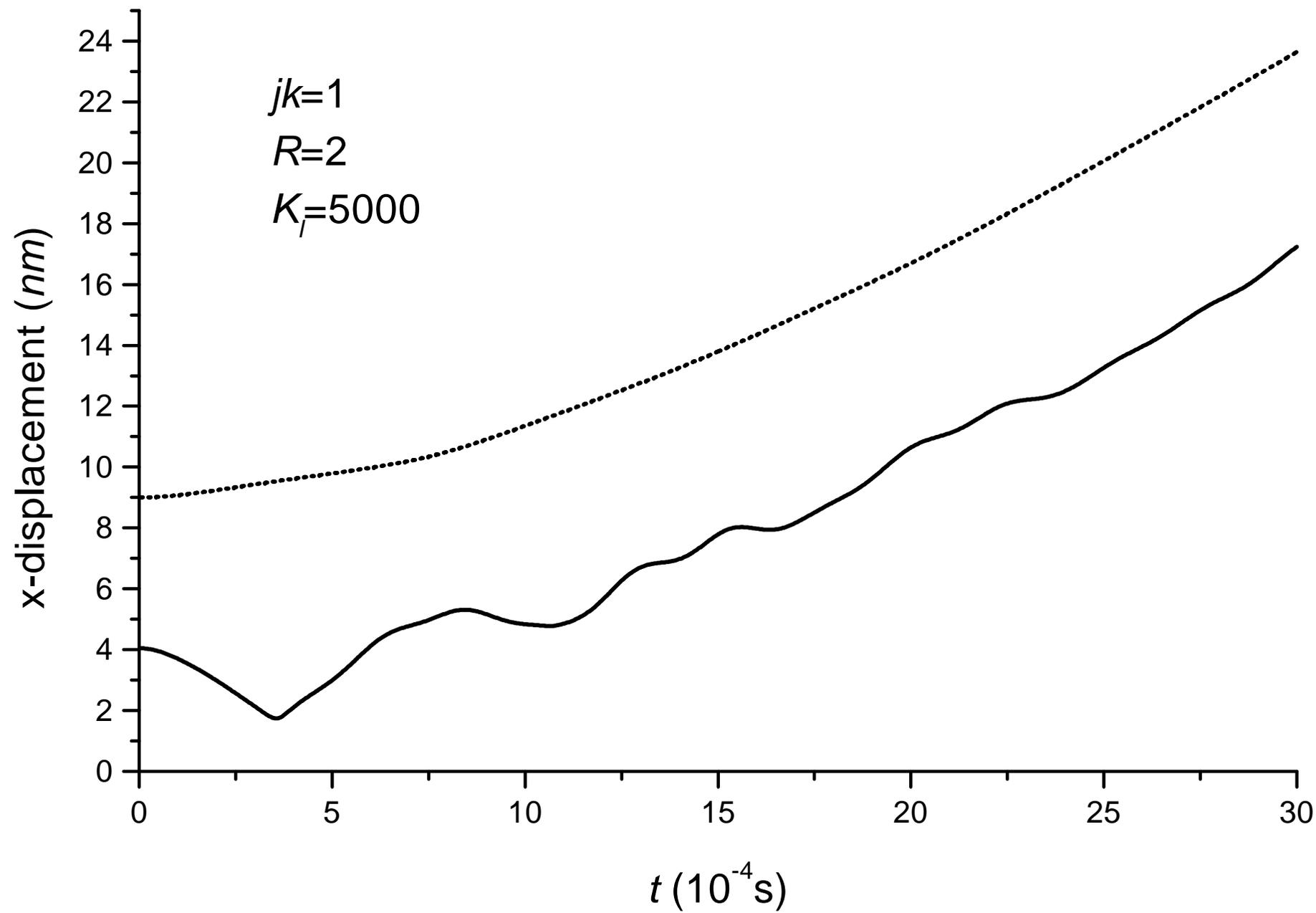
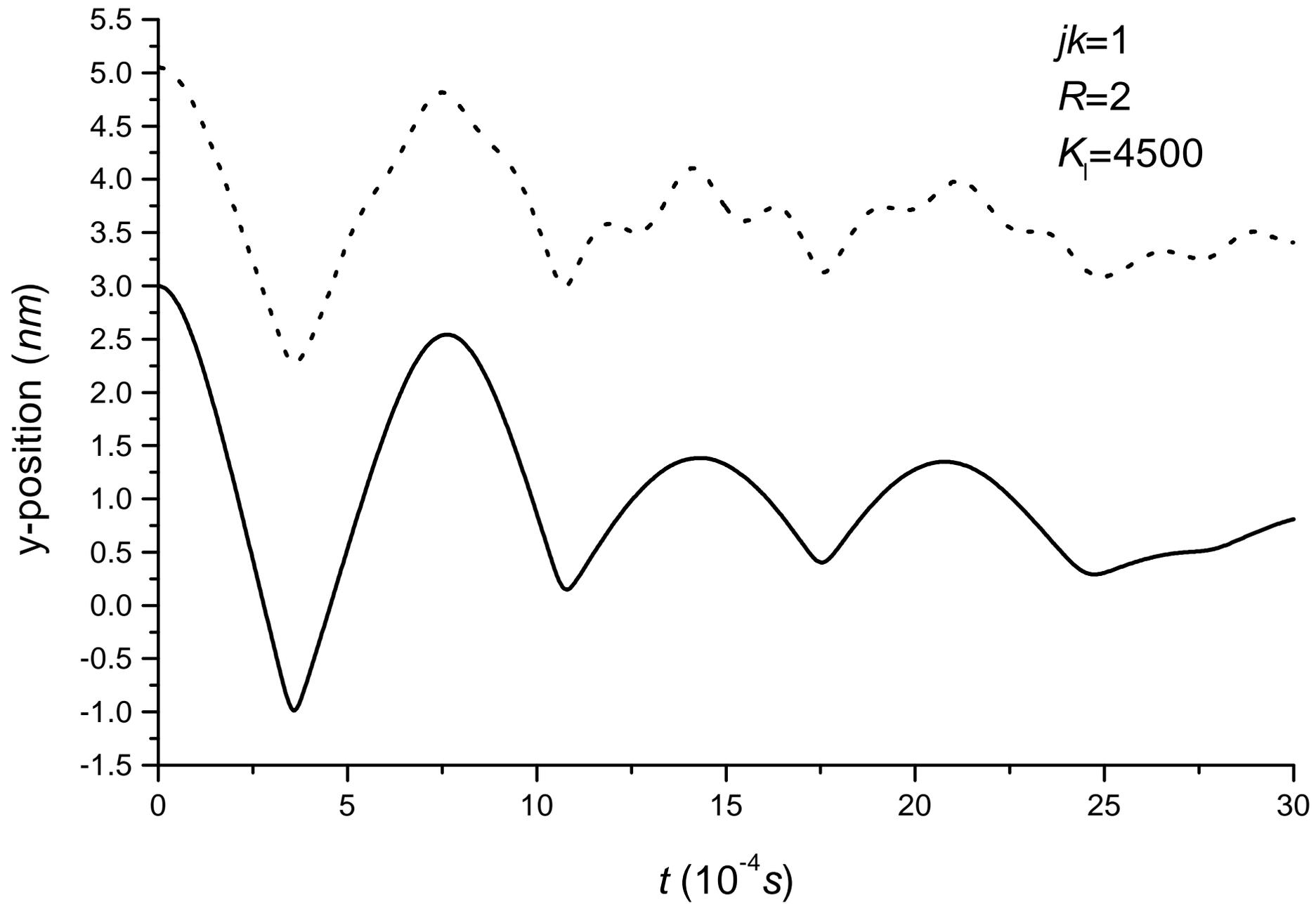
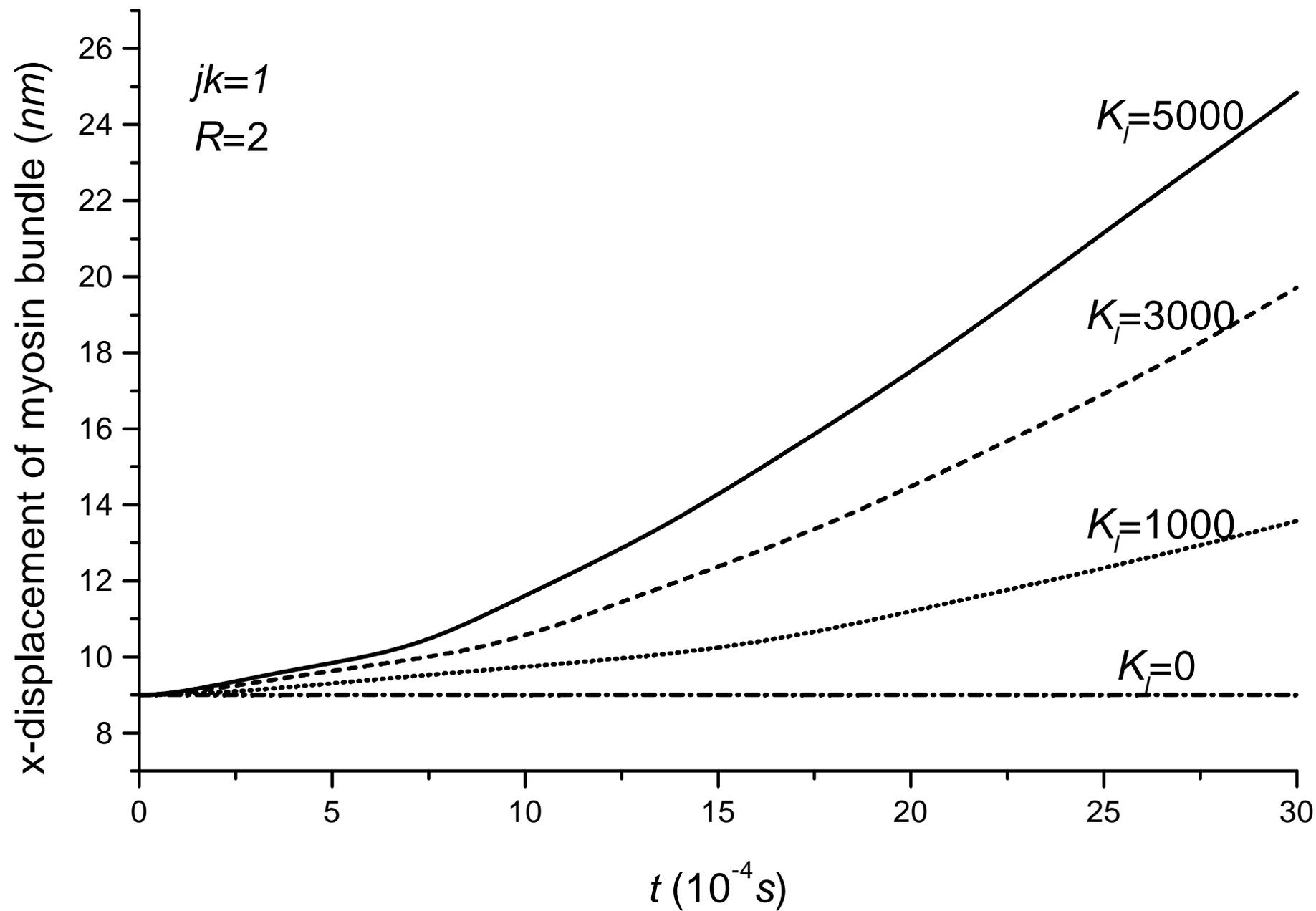
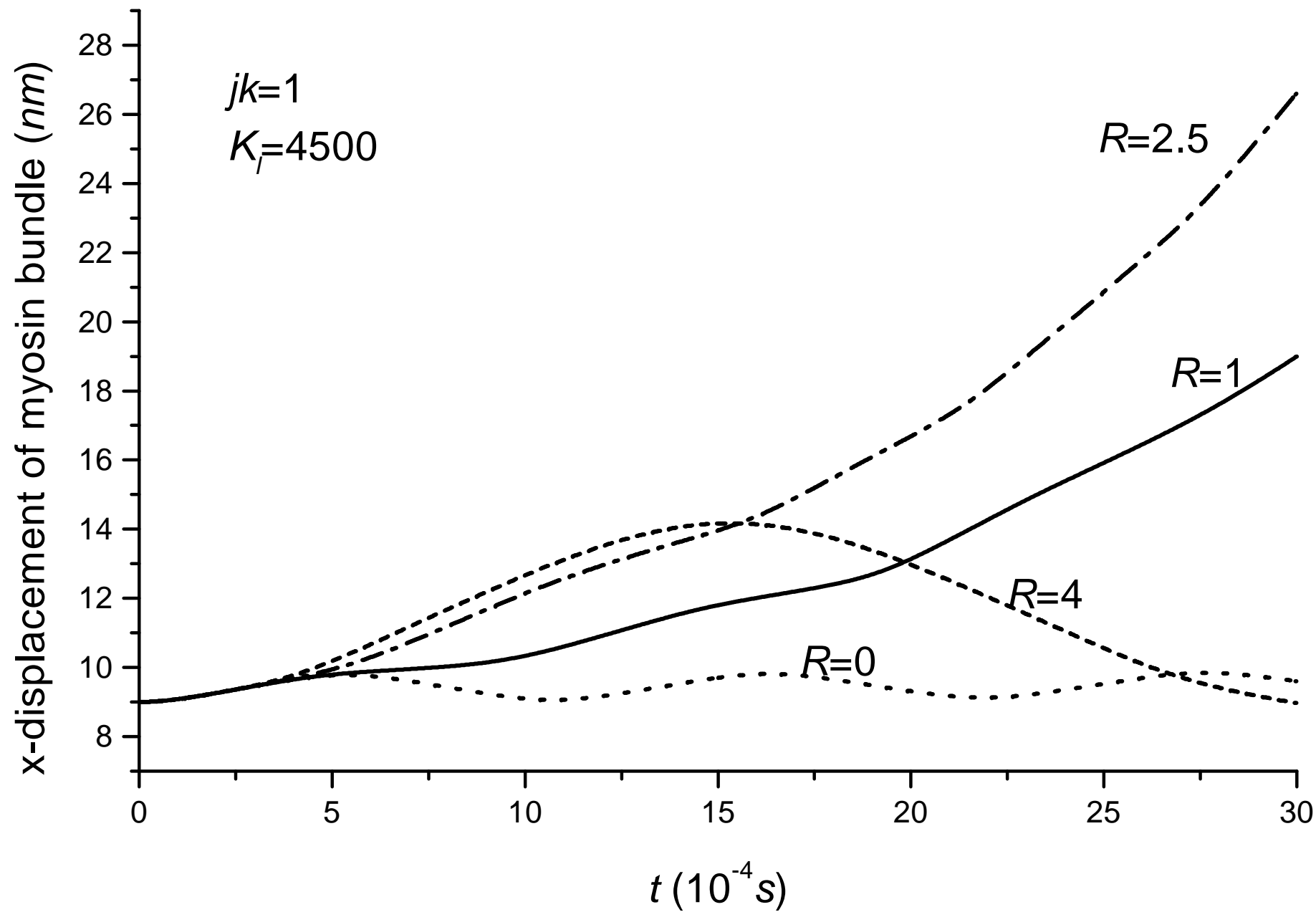
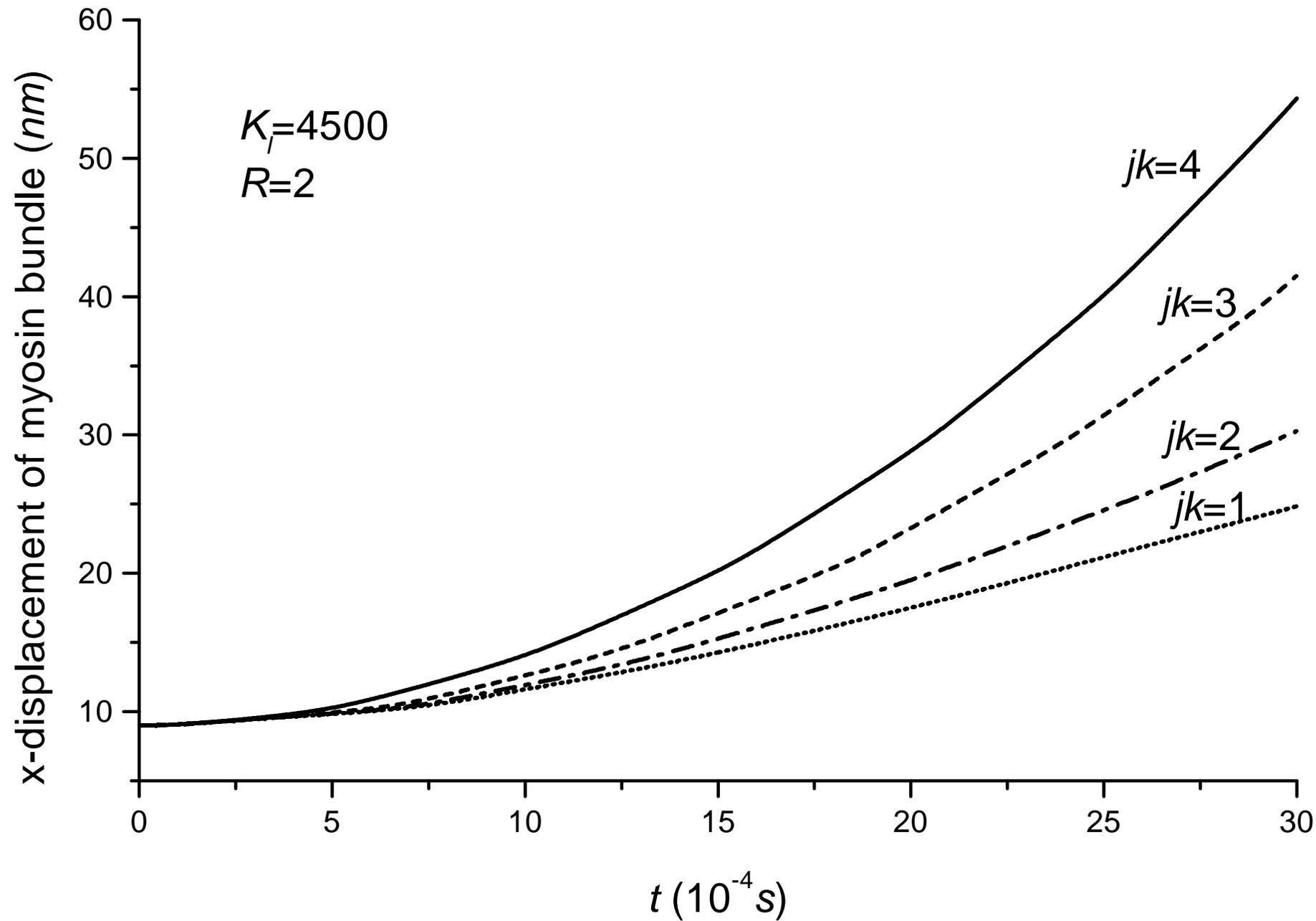
\end{document}